\documentclass[a4paper,11pt]{article}
\usepackage{pos}
\usepackage{enumitem}
\usepackage{natbib}
\renewcommand{\logo}{\relax}

\title{\textit{Euclid}: performance on main cosmological parameter science}

\author*[a,b]{Isaac Tutusaus}
\author[c,d,e]{Jenny G. Sorce}
\author[f,g]{Antonino Troja}
\author[]{on behalf of the Euclid Consortium}

\affiliation[a]{Institut de Recherche en Astrophysique et Plan\'etologie (IRAP), Universit\'e de Toulouse, CNRS, UPS, CNES, 
14 Av. Edouard Belin, F-31400 Toulouse, France}

\affiliation[b]{Universit\'e de Gen\`eve, D\'epartement de Physique Th\'eorique and Centre for Astroparticle Physics, 
24 quai Ernest-Ansermet, CH-1211 Gen\`eve 4, Switzerland}

\affiliation[c]{Univ. Lille, CNRS, Centrale Lille, UMR 9189 CRIStAL, F-59000 Lille, France}

\affiliation[d]{Universit\'e Paris-Saclay, CNRS, Institut d'Astrophysique Spatiale,
91405, Orsay, France}

\affiliation[e]{Leibniz-Institut f\"ur Astrophysik (AIP),
An der Sternwarte 16, D-14482 Potsdam, Germany}

\affiliation[f]{INFN-PD,
Via Marzolo 8, Padova, Italy}

\affiliation[g]{Universit\'a degli Studi di Padova,
Via Marzolo 8, Padova, Italy}

\emailAdd{isaac.tutusaus@irap.omp.eu}
\emailAdd{jenny.sorce@univ-lille.fr}
\emailAdd{antonino.troja@pd.infn.it}

\abstract{\textit{Euclid} will observe 15\,000 deg$^2$ of the darkest sky, in regions free of contamination by light from our Galaxy and our Solar System. Three ``Euclid Deep Fields'' surveys covering around 40 deg$^2$ in total will extend the scientific scope of the mission to the high-redshift Universe. The complete survey will be constituted by hundreds of thousands of images and several tens of petabytes of data. About 10 billion sources will be observed. With these images \textit{Euclid} will probe the expansion history of the Universe and the evolution of cosmic structures. This will be achieved by measuring the effect on galaxy shapes due to dark matter gravitational lensing, and by reconstructing the three-dimensional distribution of cosmic structures from the measured spectroscopic redshifts of galaxies and clusters of galaxies. These proceedings present the implications for cosmology and cosmological constraints of this unprecedented data set. Of particular interest are the expected constraints on the nature of dark energy.}

\FullConference{%
  41st International Conference on High Energy physics - ICHEP2022\\
  6-13 July, 2022\\
  Bologna, Italy
}

\begin{document}
\maketitle

\section{Introduction}
Within the concordance cosmological framework, dark matter tends to cluster as time evolves, giving rise to a cosmic web composed of knots, filaments, walls, and voids. Such a scenario has been tested with great accuracy using cosmological N-body simulations of dark matter particles~\cite{mice}. Galaxies, in turn, follow dark matter and tend to appear in the overdensities given by dark matter halos. Galaxy surveys provide an excellent tool to map these galaxies and directly probe both the expansion rate of the Universe and its large-scale structure (LSS). Within the concordance model, more dark matter and less dark energy leads to a larger collapse rate and, therefore, to more structures in the Universe. On the other hand, the expansion rate of the Universe is larger if there is less dark matter and more dark energy, which leads to less LSS. 

\textit{Euclid}~\cite{redbook} is a European Space Agency medium-class space mission due for launch in 2023. It will cover 15\,000 deg$^2$~\cite{Scaramella} of the extragalactic sky with the goal of measuring the geometry of the Universe and the growth of structures up to redshift 2. In order to do this, the satellite will contain two instruments on-board: a near-infrared spectro-photometer~\cite{Costille} and an imager at visible wavelengths~\cite{Cropper}. With these instruments, \textit{Euclid} will perform a photometric survey of more than a billion photometric redshifts and shapes, plus a spectroscopic survey with several tens of millions of precise spectroscopic redshift measurements. Using this extensive dataset, \textit{Euclid} will put very stringent constraints on the cosmological model driving our Universe.

These proceedings belong to a series of three proceedings together with~\cite{Antonino} and~\cite{Jenny}. The former present all the technical details of the satellite and the mission, while the latter focus on the legacy science that will be possible from the Euclid mission. These proceedings, instead, focus on presenting the performance of \textit{Euclid} on the main cosmological parameter science. The proceedings are organised as follows: In Sect.\,\ref{sec2} we briefly describe the main cosmological probes considered in \textit{Euclid} and the main constraints expected on the cosmological parameters. In Sect.\,\ref{sec3} we present some of the cosmological science that can be done with \textit{Euclid} beyond the main probes. We then present the current efforts within the Euclid Consortium to prepare the pipelines for the analysis of the real data in Sect.\,\ref{sec4}. Finally, we present our conclusions in Sect.\,\ref{sec5}.

\section{Cosmological constraints from the main probes}\label{sec2}
It is unfeasible to map the Universe LSS using the data set of galaxies, due to the large amount of information to be handled. Usually, summary statistics are used to simplify the analysis of the data. One of the main summary statistics considered in galaxy surveys is the matter (or galaxy) power spectrum. Galaxy surveys, like SDSS, BOSS, or DES, can provide extremely precise measurements at small scales, outperforming measurements from the cosmic microwave background (CMB)~\cite{Planck2018}. In the following we describe the three main cosmological probes that will be tested with the summary statistics of the spectroscopic and photometric surveys.\\

\textbf{Galaxy clustering -- spectroscopic:}
By using the precise three-dimensional galaxy positions, we can measure the so-called baryon acoustic oscillations (BAO) and the redshift-space distortions (RSDs)~\cite{eBOSS}. Once the expansion rate of the Universe was large enough, the oscillations of the primordial photon-baryon plasma froze and propagated until today. We can still see these oscillations as an excess of probability of observing galaxies at a certain separation. This separation is called the standard ruler and, by measuring it, we are sensitive to the expansion history and the angular-diameter distance.  
Concerning RSDs, we observe galaxies in redshift space, not in real space. Therefore, the redshift due to their peculiar velocities adds to the cosmological redshift. In this case, what should be an isotropic excess of probability of finding galaxies given by the BAO, becomes a squashed anisotropic distribution. By measuring RSDs we are sensitive to the growth of structures and can test the theory of gravity.\\

\textbf{Galaxy clustering -- photometric:}
\textit{Euclid} will provide a photometric survey. Although most of the radial information is lost when going from precise spectroscopic redshift measurements to photometric estimates, the number density of photometric objects will be significantly larger, which will still enable us to extract cosmological information from their clustering~\cite{ISTF}. Moreover, their systematic uncertainties will be different than those appearing in the spectroscopic sample, motivating therefore the combination of both samples in the final analysis.\\

\textbf{Weak lensing:}
Beyond considering only the position of galaxies, we can also measure their shapes. According to a relativistic theory of gravity, the geodesics followed by photons are distorted in the presence of matter. Therefore, information about the mass distribution between some sources and us as observers is imprinted on galaxy shapes. By measuring the correlation of distorted shapes of galaxies we can extract information from the matter density in the Universe, the initial conditions, and the growth of structures~\cite{Amon}.\\

\textbf{Probe combination:}
The photometric and spectroscopic surveys will overlap (at least partially) in volume. This makes \textit{Euclid} the perfect example for a combined analysis. It is well known that combining different probes can help in breaking degeneracies between different parameters and eventually lead to better parameter constraints. In \textit{Euclid} we will have three main cosmological probes in the same volume and coming from the same instruments. Therefore, the baseline analysis consists in combining the galaxy clustering (GC) measurements from both the photometric and spectroscopic samples together with the weak lensing (WL) measurements. Moreover, given the overlap in volume, additional information can be extracted from the cross-correlations between the different probes~\cite{Tutusaus}. 
Homogenised and validated forecasts were carried out in~\cite{ISTF}, concluding that:
\begin{itemize}[noitemsep,nolistsep]
    \item \textit{Euclid} will reach a dark energy figure-of-merit of 1257 (500) for a flat (non-flat) cosmology.
    \item Modified gravity (through the gravitational growth index $\gamma$) will be constrained at the $5\,\%$ level and curvature at the $1\,\%$ level.
    \item The combination of the different probes breaks several degeneracies leading to better constraints on all parameters.
    \item Cross-correlation between different probes improve constraints by a factor of 4 and are an essential ingredient to achieve the largest figure-of-merit.
\end{itemize}

\section{Beyond the main probes}\label{sec3}
Probes with cosmologically-relevant information other than the ones presented in Sect.\,\ref{sec2} can be considered, the main example being the CMB. The addition of this information at very high redshift provides us with a useful lever arm between different epochs of the Universe. An extensive analysis was performed within the Euclid Consortium to quantify the gain of combining \textit{Euclid} main probes with the lensing of the CMB~\cite{Ilic}, that is considering the CMB photons as sources and using them to probe the matter structure of the Universe. 
The full combination including the primordial CMB was also considered. Adding the CMB lensing provided by the future Simons Observatory improves the constraints on cosmological and nuisance parameters by $\sim 10$--$20\,\%$. When including all the information from the Simons Observatory measurements of the CMB, some parameters, like the baryon density, can improve up to a factor of 10. The combination of \textit{Euclid} main probes with other cosmological probes will therefore provide us with exquisite constraints on our cosmological model.

\section{Towards data analysis}\label{sec4}
There is an enormous effort within the Euclid Consortium to prepare all the pipelines for the analysis of the future data. Some open fronts are described in the following.\\

\textbf{Cosmological simulations:}
\textit{Euclid} will perform both a wide and a deep survey. Because of this, two galaxy mocks with extremely large volumes and resolutions are being prepared. These are the wide and deep Flagship galaxy catalogues, respectively. The catalogues are based on N-body dark matter simulations~\cite{Potter} where galaxies are added following a halo occupation distribution and an abundance matching technique~\cite{Carretero}. The wide Flagship galaxy mock has a resolution of $10^9\,M_{\odot}$ with 4.1 trillion particles in a 3600 Mpc$/h$ box. The deep Flagship galaxy mock, instead, has a resolution of $10^8\,M_{\odot}$ with 0.9 trillion particles in a 1000 Mpc$/h$ box.\\

\textbf{End-to-end pipeline -- spectroscopic:}
Given the complexity of the data treatment, end-to-end pipelines to go from the raw measurements down to the constraints on the cosmological parameters are being prepared. The main goal is to test that the input cosmology can be recovered, even in the presence of systematic uncertainties. Concerning the spectroscopic survey, 
the pipeline proceeds as follows: We first extract the galaxy mock catalogue from the Flagship simulation and consider astrophysical foregrounds (like the Milky Way and Zodiacal light contamination), as well as the definition of the survey patterns and dithering. With all these ingredients we generate mock raw data images, like the ones expected to be received from the instruments. These images are then processed through a reduction pipeline. After performing the measurements on the reduced images, we generate the final processed mock catalogue. We then measure our summary statistics on the catalogue and perform the cosmological inference. The latter allows us to check that we recover the input cosmology, but also that the constraining power of the survey remains as expected. Such an end-to-end pipeline is extremely useful to test the impact of different systematic uncertainties on the final cosmological results.\\

\textbf{End-to-end pipeline -- photometric:}
The size of the photometric data set is much larger, which makes a full end-to-end simulation significantly more costly. Moreover, most of the relevant systematic uncertainties for the photometric probes appear at small scales, making it less necessary to simulate all the \textit{Euclid} survey. Instead, we start with a galaxy catalogue extracted from Flagship that includes the survey dithering. Then, for each galaxy in the catalogue we assign some intrinsic shape, apply a given shear, and include the different relevant systematic uncertainties, like the point spread function, or the noise related to the detector and the observation. Since all these effects are added analytically a posteriori, we have access to both the reference and the perturbed catalogue including systematic effects. For both catalogues we can measure our summary statistics. By comparing the perturbed and unperturbed measurements, we can perform the statistical analysis and the cosmological inference. Again, such a pipeline allows us to verify that we can recover the input cosmology and that all systematic effects are under control.\\

\textbf{Next steps:}
In addition, the Euclid Consortium is building emulators to speed up the analysis and is further developing the end-to-end pipelines. There is also a significant effort on improving the modelling of the observables. In particular concerning the modelling of non-linearities of the matter and galaxy power spectra~\cite{Matteo}, or the impact of magnification and other relativistic effects~\cite{Francesca}. The pipelines are also being prepared to account for and mitigate systematic uncertainties~\cite{syst}.

\section{Conclusions}\label{sec5}
The future Euclid mission will perform an extremely high-quality spectroscopic and photometric galaxy survey that will enable us to probe the geometry of the Universe and the growth of structures up to high redshift. By combining the main cosmological probes (GC and WL), \textit{Euclid} alone will constrain the equation of state parameter of dark energy at the level of $3\,\%$. This corresponds to an improvement by a factor of 3 with respect to current constraints~\cite{desext}, which include photometric GC, WL, RSDs, BAO, type-Ia supernovae, and CMB measurements. In conclusion, \textit{Euclid} will provide us with unprecedented constraints on dark matter and gravity at cosmological scales, and will constrain dark energy better than all current observations together.\\

{\small \textbf{Acknowledgments:}
 IT acknowledges funding from the Swiss National Science Foundation and from the European Research Council (ERC) under the European Union's Horizon 2020 research and innovation program (Grant agreement No.~863929). JS acknowledges support from the French Agence Nationale de la Recherche for the LOCALIZATION project under grant agreements ANR-21-CE31-0019. The Euclid Consortium acknowledges the European
  Space Agency and a number of agencies and
  institutes that have supported the development of \textit{Euclid}, in
  particular the Academy of Finland, the Agenzia Spaziale
  Italiana, the Belgian Science Policy, the Canadian Euclid
  Consortium, the French Centre National d'Etudes Spatiales, the
  Deutsches Zentrum f\"ur Luft- und Raumfahrt, the Danish Space
  Research Institute, the Funda\c{c}\~{a}o para a Ci\^{e}ncia e a
  Tecnologia, the Ministerio de Ciencia e
  Innovaci\'{o}n, the National Aeronautics and Space
  Administration, the National Astronomical Observatory of Japan,
  the Netherlandse Onderzoekschool Voor Astronomie, the Norwegian
  Space Agency, the Romanian Space Agency, the State Secretariat
  for Education, Research and Innovation (SERI) at the Swiss
  Space Office (SSO), and the United Kingdom Space Agency. A
  complete and detailed list is available on the \textit{Euclid} web site
  (\texttt{http://www.euclid-ec.org}).}

\end{document}